\global\def\draftcontrol{0}

   \def\versionno{transport at criticality }

\catcode`\@=11

\expandafter\ifx\csname draftcontrol\endcsname\relax\global\def\draftcontrol{0}
\fi

{\count255=\time\divide\count255 by 60
\xdef\hourmin{\number\count255}
\multiply\count255 by-60\advance\count255 by\time
\xdef\hourmin{\hourmin:\ifnum\count255<10 0\fi\the\count255}}
\def\draftdate{\number\month/\number\day/\number\year\ \ \ \hourmin }

\newcommand\makepapertitle{\par
  \begingroup
    \renewcommand\thefootnote{\@fnsymbol\c@footnote}%
    \def\@makefnmark{\rlap{\@textsuperscript{\normalfont\@thefnmark}}}%
    \long\def\@makefntext##1{\parindent 1em\noindent
            \hb@xt@1.8em{%
                \hss\@textsuperscript{\normalfont\@thefnmark}}##1}%
     \newpage
     \global\@topnum\z@   
     \@makepapertitle
     \thispagestyle{empty}\@thanks
  \endgroup
  \setcounter{footnote}{0}%
  \global\let\thanks\relax
  \global\let\makepapertitle\relax
  \global\let\@makepapertitle\relax
  \global\let\@thanks\@empty
  \global\let\@author\@empty
  \global\let\@date\@empty
  \global\let\@title\@empty
  \global\let\title\relax
  \global\let\author\relax
  \global\let\date\relax
  \global\let\and\relax
  \def\version{\let\version\@version\@gobble}
}
\def\@makepapertitle{%
  \newpage
   \ifnum\draftcontrol=1 {}
   \version\versionno
   \vskip 3em%
   \else
   \hfill\hbox to 3cm {\parbox{4cm}{\@pubnum}\hss}%
   \vskip 3em%
   \fi
   \begin{center}%
   \let \footnote \thanks
     {\LARGE {\@title}}%
     \vskip 1.5em%
     {\normalsize
       \lineskip .5em%
       \begin{tabular}[t]{c}%
         \@author
       \end{tabular}\par}%
     \vskip 1.5em%
     {\@bstract}%
     \end{center}%
     \vskip 1.5em
     \@date%
   \par
}

\gdef\@pubnum{}
\def\pubnum#1{%
  \gdef\@pubnum{#1}}

\gdef\@bstract{}
\def\Abstract#1{%
  \gdef\@bstract{%
   \parbox{\textwidth-0pc}{%
   \centerline{\bf Abstract}\penalty1000%
\kern.2cm%
\noindent
\renewcommand\baselinestretch{1.0}%
{#1}}}
}

\def\ps@paper{\let\@mkboth\@gobbletwo%
     \ifnum\draftcontrol=1
    \def\@oddfoot{\hbox to \textwidth{\tiny \versionno \hfil\tiny\draftdate}%
    \hskip -\textwidth \hbox to \textwidth{\hfil\rm\thepage\hfil}}%
     \else\def\@oddfoot{\hbox to \textwidth{\hfil\rm\thepage\hfil}}
     \fi
     \let\@evenfoot\@oddfoot
}

\def\body{\clearpage
          \pagestyle{paper}
    }

\def\@version#1{\ifnum\draftcontrol=1
\typeout{}\typeout{#1}\typeout{}
\vskip3mm\centerline{\hbox{\fbox{\normalsize{\tt DRAFT -- #1 -- }
                   {\draftdate}}}}\vskip3mm
\fi}
\let\version\@version
\long\def\eqlabel#1{\ifnum\draftcontrol=1
                    \tag@false  
                    \tag*{(\theequation) \hbox to -0.2cm{\hspace{0cm}\small{#1}\hss}}
                    \refstepcounter{equation}
                    \edef\@currentlabel{\theequation}
                    \ltx@label{#1}          
                    \else
                    \label{#1}
                    \fi
                    }
\let\st@bibitem\@bibitem
\let\st@lbibitem\@lbibitem
\ifnum\draftcontrol=1
  \def\@bibitem#1{%
    \st@bibitem{#1}\a@@label{#1}\ignorespaces}
  \def\@lbibitem[#1]#2{%
    \st@lbibitem[#1]{#2}\a@@label{#2}\ignorespaces}
  \def\a@@label#1{%
    \gdef\a@lab{\smash{\normalfont\small#1}}
    \ifvmode
      \if@inlabel
        \global\setbox\@labels\hbox{%
          \llap{\a@lab\let\a@lab\relax
                \kern\@totalleftmargin\kern\marginparsep}%
          \box\@labels}%
      \fi
    \fi}
\fi

\documentclass[12pt,letterpaper]{article}

\usepackage{amsmath,amssymb,array,calc,epsfig}
\usepackage{graphicx}
\usepackage{psfrag,verbatim,bm}

\ifnum\draftcontrol=1
\fi

\renewcommand\baselinestretch{1.25}
\renewcommand\section{\@startsection {section}{1}{\z@}%
                                   {-3.5ex \@plus -1ex \@minus -.2ex}%
                                   {2.3ex \@plus.2ex}%
                                   {\normalfont\large\bfseries}}
\renewcommand\subsection{\@startsection{subsection}{2}{\z@}%
                                   {-3.25ex\@plus -1ex \@minus -.2ex}%
                                   {1.5ex \@plus .2ex}%
                                   {\normalfont\normalsize\bfseries}}
\renewcommand\subsubsection{\@startsection{subsubsection}{3}{\z@}%
                                   {-3.25ex\@plus -1ex \@minus -.2ex}%
                                   {1.5ex \@plus .2ex}%
                                   {\normalfont\normalsize\it}}
\renewcommand\paragraph{\@startsection{paragraph}{4}{\z@}%
                                   {-3.25ex\@plus -1ex \@minus -.2ex}%
                                   {1.5ex \@plus .2ex}%
                                   {\normalfont\normalsize\bf}}

\numberwithin{equation}{section}

\def\ie{{\it i.e.}}

\def\revise#1       {\raisebox{-0em}{\rule{3pt}{1em}}%
                     \marginpar{\raisebox{.5em}{\vrule width3pt\
                     \vrule width0pt height 0pt depth0.5em
                     \hbox to 0cm{\hspace{0cm}{%
                     \parbox[t]{4em}{\raggedright\footnotesize{#1}}}\hss}}}}

\newcommand\nxt[1]  {\\\fnxt#1}

\def\calh         {{\cal H}}

\def\calm         {{\cal M}}
\def\caln         {{\cal N}}
\def\calo         {{\cal O}}

\def\calw         {{\cal W}}

\def\zet          {{\mathbb Z}}

\def\del          {\partial}

\def\sqr#1#2{{\vcenter{\vbox{\hrule height.#2pt
 \hbox{\vrule width.#2pt height#1pt \kern#1pt
 \vrule width.#2pt}\hrule height.#2pt}}}}

\def\a{\alpha}
\def\b{\beta}
\def\l{\lambda}

\def\r{\rho}

\newcommand{\qq}{\mathfrak{q}}
\newcommand{\ww}{\mathfrak{w}}

\def\e{\epsilon}

\catcode`\@=12

\begin{document}


\title{\bf Transport at criticality}
\pubnum
{UWO-TH-09/17
}

\date{December 2009}

\author{
Alex Buchel$ ^{1,2}$ and Chris Pagnutti$ ^{1}$\\[0.4cm]
\it $ ^1$Department of Applied Mathematics\\
\it University of Western Ontario\\
\it London, Ontario N6A 5B7, Canada\\
\it $ ^2$Perimeter Institute for Theoretical Physics\\
\it Waterloo, Ontario N2J 2W9, Canada\\
}

\Abstract{
We study second order phase transitions in non-conformal holographic
models of gauge theory/string theory correspondence at finite
temperature and zero chemical potential. We compute critical exponents
of the bulk viscosity near the transition and interpret our results in
the framework of available models of dynamical critical
phenomena. Intriguingly, although some of the models we discuss belong
to different static universality classes, they appear to share the
same dynamical critical exponent.
}

\makepapertitle

\body

\version\versionno

\section{Introduction}
Gauge theory/string theory correspondence of Maldacena \cite{m9711} presents a useful theoretical 
framework to study strong coupling dynamics of gauge theories. In this paper we use it to study 
bulk viscosity of  non-conformal gauge theory plasma at criticality of a second 
order phase transition. We focus on  finite temperature phase transitions 
at zero chemical potential\footnote{Critical phenomena in strongly coupled plasma
at finite density  will be discussed in \cite{bpta}. }. Furthermore, since the models we consider 
do not have  long-range (Goldstone boson) modes at the criticality, their hydrodynamics is 
that of a standard viscous relativistic fluid\,, i.e, the local stress-energy tensor 
is given by 
\begin{equation}
\begin{split}
T^{\mu\nu}=&\e u^{\mu}u^{\nu}+P(\e)\Delta^{\mu\nu}-\eta(\e)\sigma^{\mu\nu}-\zeta(\e)\Delta^{\mu\nu}
\left(\nabla\cdot u\right)\,,\\
\Delta^{\mu\nu}=&g^{\mu\nu}+u^\mu u^\nu\,,\qquad 
\sigma^{\mu\nu}=\Delta^{\mu\a}\Delta^{\nu\b}(\nabla_{\a}u_\b+\nabla_\b u_\a)-\frac{2}{d-1}\Delta^{\mu\nu}
\Delta^{\a\b}\nabla_\a u_\b\,,
\end{split}
\eqlabel{tmn}
\end{equation}
where $\e$ and $P(\e)$ are the local energy density and pressure, $u^\mu$ is the local $d$-velocity
of the plasma, and  $\eta(\e)$ and $\zeta(\e)$ are the shear and the bulk viscosities 
correspondingly. A plasma with the stress-energy tensor \eqref{tmn} allows for a propagation of the 
hydrodynamic sound waves with the following dispersion relation
\begin{equation}
\ww=c_s^2\ \qq-i\ \Gamma\ \qq^2+\calo\left(\qq^3\right)\,,
\eqlabel{sd}
\end{equation} 
where $c_s$ is the speed of sound and $\Gamma$ is the sound wave attenuation,
\begin{equation}
c_s^2=\left(\frac{\del P}{\del \e}\right)_T=\frac{s}{c_v}\,,\qquad \Gamma=2\pi\ \frac{\eta}{s}\left(
\frac{d-2}{d-1}+\frac{\zeta}{2\eta}
\right)\,,
\eqlabel{csG}
\end{equation} 
and 
$\ww=\omega/(2\pi T)$ and $\qq=|\vec{q}|/(2\pi T)$. Thus, if the dispersion relation for the 
sound waves \eqref{sd} in plasma can be determined from first principles, the bulk viscosity 
can be computed from \eqref{csG}\footnote{One also needs a ratio of shear viscosity to the entropy 
density. In  models we analyze this ratio is universal \cite{u1,u2,u3}.}. 
A holographic correspondence of Maldacena provides a framework 
for such first principle computations \cite{kovs}: the dispersion relation of the sound waves in
strongly coupled plasma is identified with the dispersion relation of a hydrodynamic 
graviton quasinormal mode (of a certain polarization) in gravitational background holographically dual 
to the equilibrium thermal state of the plasma. 

In a conformal plasma the trace of the stress-energy tensor vanishes. As a result, 
\begin{equation}
c_s^2\bigg|_{CFT}=\frac{1}{d-1}\,,\qquad \frac{\zeta}{\eta}\bigg|_{CFT}=0\,. 
\eqlabel{cft}   
\end{equation}
However, once the scale invariance is explicitly broken, the bulk viscosity is generically nonzero.
While for most physical substances  the bulk viscosity is smaller or at most of the same order 
as the shear viscosity, $ ^3$He in the vicinity of the critical liquid-vapor point exhibits the ratio 
of the bulk-to-shear viscosities in the excess of a million \cite{kh}. 
Below, we present the first example of a holographic model with divergent ratio of $\frac{\zeta}{\eta}$
in the vicinity of a second order phase transition. 
   
Techniques for the holographic computations of the bulk viscosity in non-conformal plasma 
were developed in \cite{bbs}. In what follows we omit all the technical details of the analysis.
In the next section we mention three different proposals for the scaling of bulk viscosity 
at the criticality of a second order phase transition \cite{bulk1,bulk2,bulk3}. 
In section 3, building up on the previous work \cite{n2tr1,n2tr2,castr1},
we identify the second order phase transitions (and the static universality classes) 
in holographic models of gauge-gravity correspondence:
the  $\caln=2^*$ gauge theory plasma \cite{pw1,pw2,pw3}, the finite temperature Klebanov-Tseytlin (cascading) gauge theory
plasma \cite{ca1,ca2,ca3,ca4,ca5,ca6}, and the phase transition in "exotic black holes'' we proposed earlier \cite{bp1}.
In section 4 we present results for the scaling of bulk viscosity at the criticality in these models. 
While the computations of bulk viscosity in $\caln=2^*$ and the cascading gauge theories were done previously 
(see \cite{bbs,n2tr2,bp2,castr0,castr1} ), 
our result for the bulk viscosity of plasma dual to "exotic black holes'' of \cite{bp1} is new.    
We conclude in section 5 by interpreting the scaling of the bulk viscosity in holographic models  in the framework
of models of dynamical critical phenomena \cite{bulk1,bulk2,bulk3}. 

\section{Models of bulk viscosity at criticality}   
We end up mapping the  second order phase transitions in holographic models  of gauge/gravity correspondence to the 
ferromagnetic phase transition in $p=d-1$ spatial dimensions --- thus we use the nomenclature of the latter. 

Consider the Gibbs free energy density $\calw=\calw(T,\calh)$ which is the difference between the free energy densities 
of the ordered and the disordered phase 
\begin{equation}
\calw(T,\calh)=\Omega_{o}(T,\calh)-\Omega_d(T,\calh)
\eqlabel{diffw}
\end{equation}
as a function of temperature $T$ and the (generalized) external magnetic field $\calh$. The (generalized) spontaneous magnetization 
$\calm$ determines the response of the free energy to the changes in the external control parameter, $\calh$, as 
\begin{equation}
\calm=-\left(\frac {\del\calw}{\del\calh}\right)_T\,.
\eqlabel{mag}
\end{equation}
The energy density $\e$, the entropy density $s$, and the spontaneous magnetization $\calm$ satisfy the basic thermodynamic 
relation
\begin{equation}
\calw=\e-s\ T - \calm\ \calh\,,
\eqlabel{basic}
\end{equation}
with the first law of thermodynamics
\begin{equation}
d\calw=-s\ dT-\calm\ d\calh\,.
\eqlabel{fl}
\end{equation}
For convenience, we introduce a reduced temperature $t$ 
\begin{equation}
t\equiv \frac{T-T_c}{T_c}\,,
\eqlabel{t}
\end{equation}
where $T_c$ is the critical temperature of the second order  phase transition.
For a standard ferromagnetic phase transition\footnote{Some of the holographic 
phase transitions we discuss are ``not standard'', see \cite{bp1}.}
\begin{equation}
\begin{split}
&t<0:\qquad \calw(t,\calh)<0\,,\qquad \calm(t,\calh=0)\ne 0\,,\\
&t>0:\qquad \calw(t,\calh)>0\,,\qquad \calm(t,\calh=0)=0\,.
\end{split}
\eqlabel{ferro}
\end{equation}
At a second order phase transition the first derivatives of $\calw$ are continuous while the higher derivatives 
are not. Under the static scaling hypothesis we have
\begin{equation}
\calw(t,\calh)=\lambda^{-p}\ \calw\left(\lambda^{y_T} t,\lambda^{y_\calh} \calh\right)\,,
\eqlabel{staticscaling1}
\end{equation}
for the free energy, and 
\begin{equation}
\tilde{G}(\vec{q},t,\calh)=\l^{2 y_\calh-p}\ \tilde{G}(\l \vec{q},\l^{y_T} t,\l^{y_\calh} 
\calh)\,,
\eqlabel{staticscaling2}
\end{equation}
for the Fourier transform of the equilibrium two-point correlation function of the
magnetization
\begin{equation}
G(\vec{r})=\langle \calm(\vec{r})\calm(\vec 0)\rangle\propto \frac{\del^2\calw}{\del\calh(\vec{r})\del\calh(\vec 0)}\,.
\eqlabel{defg}
\end{equation} 

The static critical exponents $\{\a,\b,\gamma,\delta,\nu,\eta\}$ are defined as 
\begin{equation}
\begin{split}
&c_\calh=-T\left(\frac{\del^2 \calw}{\del T^2}\right)_\calh\propto |t|^{-\a}\,,\qquad \calm\propto |t|^\b\,,\qquad\\
&\chi_T=\left(\frac{\del \calm}{\del\calh}\right)_T\propto |t|^{-\gamma}\,,\qquad \calm(t=0)\propto |\calh|^{1/\delta} \,,
\end{split}
\eqlabel{defs0}
\end{equation}
and 
\begin{equation}
G(\vec{r})\propto
\begin{cases}
&e^{-|\vec{r}|/\xi}\,,\qquad  t\ne 0\cr
&|\vec{r}|^{-p+2-\eta}\,,\qquad  t=0
\end{cases}\,,\qquad {\rm with}\qquad \xi\propto |t|^{-\nu} \,,
\eqlabel{defs01}
\end{equation}
where $c_\calh$ is the specific heat, $\chi_T$ is the magnetic susceptibility, and $\xi$ is the correlation length.
Given \eqref{fl} and the scaling hypothesis \eqref{staticscaling1} and \eqref{staticscaling2}, 
the two critical exponents $\{y_T,y_\calh\}$ determine the  critical exponents \eqref{defs0} 
which identify the standard static universality classes:
\begin{equation}
\begin{split}
&\a=2-\frac{p}{y_T}\,,\qquad \b=\frac{p-y_\calh}{y_T}\,,\qquad \gamma=\frac{2y_\calh-p}{y_T}\,,\\
&\delta=\frac{y_\calh}{p-y_\calh}\,,\qquad \nu=\frac{1}{y_T}\,,\qquad \eta=p-2y_\calh+2\,.
\end{split}
\eqlabel{defs}
\end{equation}
Note that \eqref{defs} implies the following scaling relations
\begin{equation}
\begin{split}
\a+2\b +\gamma=2\,,\qquad \gamma=\b(\delta-1)=\nu(2-\eta)\,,\qquad 2-\a=\nu p\,.
\end{split}
\eqlabel{screl}
\end{equation}
The scaling relations \eqref{screl} can be violated whenever the single-scale hypotheses \eqref{staticscaling1} and 
\eqref{staticscaling2} break down. In what follows we assume that this is not the case in our models.

A finer classification of the universality classes of systems undergoing second order phase transition occurs 
once the dynamical properties of the system (such as hydrodynamics) are under consideration \cite{hh}.  
Here, the main assumption is that a system perturbed away from the equilibrium will relax with a characteristic 
time scale $\tau_q$, which diverges near the phase transition with a dynamical critical exponent $z$:
\begin{equation}
\tau_q\propto \xi^z\propto |t|^{-\nu z }\,.
\eqlabel{taus}
\end{equation}
Correspondingly, the near-equilibrium scaling of the magnetization correlation function \eqref{staticscaling2} is modified 
according to 
\begin{equation}
\tilde{G}(\omega,\vec{q},t,\calh)=\l^{2 y_\calh-d+z}\ \tilde{G}(\l^z\omega,\l \vec{q},\l^{y_T} t,\l^{y_\calh} \calh)\,.
\eqlabel{staticscaling2n}
\end{equation}

In this paper we will be interested in the hydrodynamic aspects of the critical phenomena, specifically in the scaling of the 
bulk viscosity at the transition. Although second order transitions imply {\it spatial} scale invariance \eqref{defs01}, 
the latter does not preclude a non-zero bulk viscosity which necessitates the {\it space-time} scale invariance, 
see \eqref{cft}. A large bulk viscosity at the transition signals strong coupling between the dilatational modes of the 
system and its internal degrees of freedom. This is a very interesting phenomena given that hydrodynamics is often 
regarded as an effective low-energy description of a system. Without going into details we now mention three proposals (Models A,B,C)
for the scaling of the bulk viscosity at criticality\footnote{Here we indicate only the scaling of the potentially 
singular part of the bulk viscosity. A finite contribution is always implicit.}.
\nxt (Model A):\ In \cite{bulk1} the authors proposed that the critical behavior of the bulk viscosity is governed by the critical 
exponent of the specific heat 
\begin{equation}
\zeta\propto c_\calh \propto |t|^{-\a}\,.
\eqlabel{kkts}
\end{equation}   
\nxt (Model B):\ In the quasi-particle model under the relaxation time approximation the bulk viscosity was argued to scale as \cite{bulk2}
\begin{equation}
\zeta\propto |t|^{\a+4\b-1}\,.
\eqlabel{sr}
\end{equation}
\nxt (Model C):\ In \cite{bulk3} Onuki argued that the bulk viscosity near the gas-liquid critical point 
(model H in classification of \cite{hh}) scales as 
\begin{equation}
\zeta\propto |t|^{-z\nu+\a}\,.
\eqlabel{on}
\end{equation}

Below, we discuss second order phase transitions in holographic models of gauge/gravity correspondence. 
We show that only the scaling \eqref{on} passes these holographic tests.  

\section{Holographic second order phase transitions}
A second order phase transition in a holographic model of gauge/gravity correspondence was first reported in 
\cite{hhh}. This phase transition is associated with the spontaneous breaking of $U(1)$ symmetry, and as a result 
the hydrodynamics of the model must include a long-range Goldstone boson mode. Another type of a holographic 
second order phase transition was discussed in \cite{maeda}.  The phase  transitions in \cite{hhh} and \cite{maeda}
occur both at finite temperature and finite chemical potential. 

We discuss bulk viscosity in second order phase transitions at finite density in \cite{bpta}. Here, we focus on 
phase transitions at vanishing chemical potential for the conserved charges. An example of such transition
was reported in \cite{bp1} (we review the main features of this transition in section 3.3). We show that both the $\caln=2^*$
gauge theory and the cascading gauge theory plasma undergo the second order phase transitions which are in the same 
static universality class. The latter universality class includes the model of \cite{maeda}, but is different from the 
universality class of the model \cite{bp1}. The static universality classes of the holographic models we discuss below are not 
of the mean-field type since the critical exponent $\eta$  (see \eqref{defs01}) is nonzero.

\subsection{$N=2^*$ plasma }

\begin{figure}[t]
\begin{center}
\psfrag{cs2}{{$c_s^2$}}
\psfrag{r11}{{$\r_{11}$}}
\psfrag{mbT}{{$\frac{m_b}{T}$}}
  \includegraphics[width=3in]{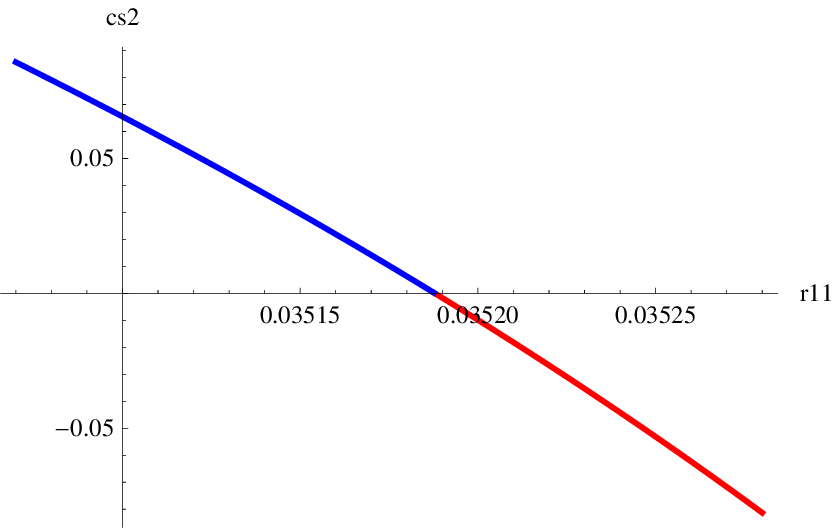}
\includegraphics[width=3in]{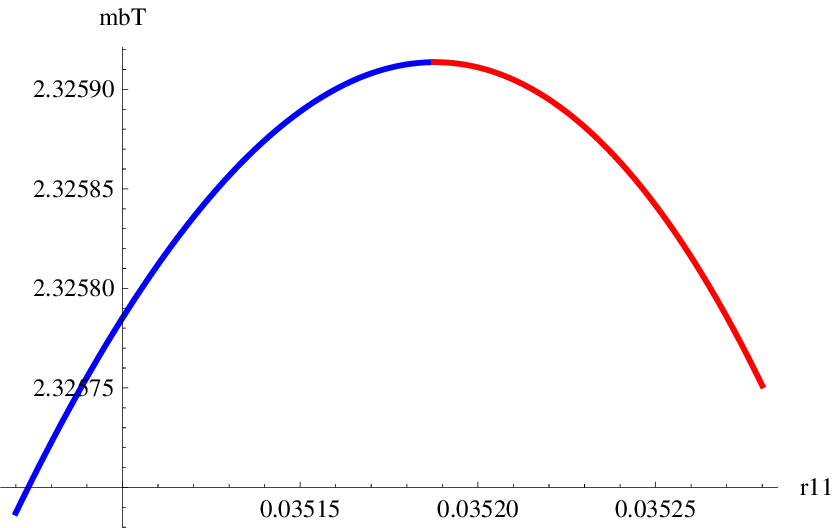}
\end{center}
  \caption{(Colour online)
The speed of sound $c_s^2$ (left plot) and the reduced temperature $\frac{m_b}{T}$ (right plot)
of the strongly coupled $\caln=2^*$ plasma with $m_f=0$ and $m_b\ne 0$
as a function of the dual gravitation parameter $\r_{11}$. We identify the blue curve 
with the "ordered'' phase and the red curve with the "disordered'' phase, see \eqref{diffw}.} \label{figure1}
\end{figure}

\begin{figure}[t]
\begin{center}
\psfrag{fred}{{$\Omega/\left(\frac 18 \pi^2 N^2 T^4\right)$}}
\psfrag{r11}{{$\r_{11}$}}
\psfrag{mbT}{{$\frac{m_b}{T}$}}
  \includegraphics[width=3in]{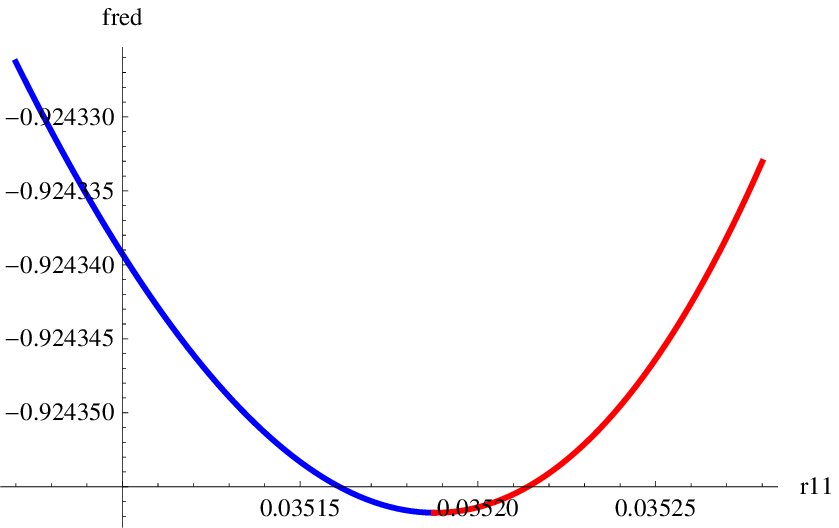}
\includegraphics[width=3in]{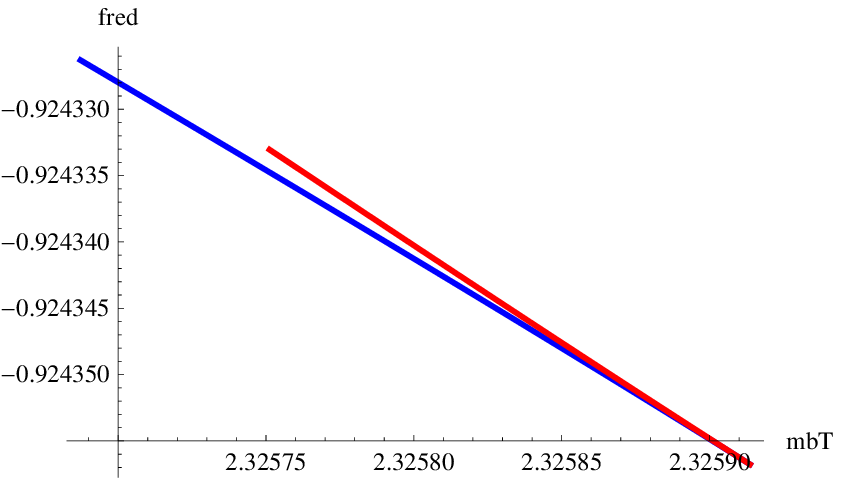}
\end{center}
  \caption{(Colour online)
Free energy densities $\Omega_o$ of the ``ordered'' phase (blue curves) 
and $\Omega_d$ of the ``disordered'' phase (red curves) as a function of 
$\r_{11}$ (left plot) and $\frac{m_b}{T}$ (right plot) of the $\caln=2^*$
plasma with $m_f=0$.} \label{figure2}
\end{figure}

The holographic renormalization and the thermodynamics of the $\caln=2^*$ gauge theory plasma in the planar limit
and at (infinitely) large 't Hooft coupling was discussed previously
\cite{n2e1,n2e2,n2tr1}\footnote{Thermodynamics of weakly coupled $\caln=2$ $SU(2)$ Yang-Mills theory was recently discussed in 
\cite{larry}.
}. This gauge theory is obtained by giving a mass to one of the three $\caln=2$ hypermultiplets of 
$\caln=4$ $SU(N)$ supersymmetric Yang-Mills theory. At finite temperature the supersymmetry is broken, and one can study 
the theory with different masses for bosonic and fermionic components of the massive hypermultiplet.
In \cite{n2tr1} the different mass deformations were called the ``bosonic'' $m_b$,  and the ``fermionic'' $m_f$.
For the range of parameters $(T,m_b,m_f)$ studied in \cite{n2tr1,n2tr2} the $\caln=2^*$ plasma is always in 
deconfined phase\footnote{An interesting phase transition in this model was conjectured in \cite{n2e1}, see also
\cite{larry}. }. It was further shown that whenever $m_f^2< m_b^2$ the theory undergoes a phase transition with 
the vanishing speed of sound. Although the critical temperature $T_c$ of the transition depends on the ratio 
$\frac{m_f^2}{m_b^2}$, the general thermodynamic and the hydrodynamic features of the model are the same \cite{bp2}.   
Thus, without the loss of generality we consider the mass deformation $m_f=0$ and $m_b\ne 0$. Here, the transition occurs 
at \cite{n2tr1} 
\begin{equation}
\frac{m_b}{T_c}\approx 2.32591\,.
\eqlabel{tcn2}
\end{equation}
We now identify this phase transition as a second order phase transition with the static critical 
exponents 
\begin{equation}
\left(\a,\b,\gamma,\delta,\nu,\eta\right)=\left(\frac 12,\frac 12,\frac 12, 2,\frac 12, 1 \right)\,.
\eqlabel{staticn2}
\end{equation}
 
It is convenient to present the thermodynamic data as a function of the dual gravitational parameter $\r_{11}$. 
For $T\gg m_b$ we have \cite{n2e1} 
\begin{equation}
\r_{11}=\frac{\sqrt{2}}{24\pi^2}\ \left(\frac{m_b}{T}\right)^2 +\calo\left(\frac{m_b^4}{T^4}\right)\,,
\eqlabel{defr11}
\end{equation} 
for all other temperatures see section 3.3 of \cite{n2tr1}. The left plot on Fig.~\ref{figure1} represents the square of the 
speed of sound $c_s^2$ as a function of $\r_{11}$ while the right plot  on Fig.~\ref{figure1} represents the reduced temperature 
$\frac{m_b}{T}$ as a function of $\r_{11}$. The critical --- the lowest --- temperature corresponds to the maximum on the plot.  
The plots on Fig.~\ref{figure2} represent the (reduced) free energy of the system $\Omega$ as a function of $\r_{11}$ (left plot)
and the reduced temperature (right plot).   
For a given temperature $T>T_c$ there are two phases of $\caln=2^*$ plasma denoted by a blue/red dot. 
From Fig.~\ref{figure2} it is clear that the transition between the two phases is a continuous one.
The ``red''
phase is perturbatively unstable as it has $c_s^2\bigg|_{red}<0$ --- it is natural to identify it with the disordered phase of the 
effective ferromagnet in section 2. Perturbatively stable  "blue'' phase is thus an ordered one. From the right plot 
in Fig.~\ref{figure2} we see that the ordered phase is thermodynamically favorable since
\begin{equation}
\calw=\Omega_o-\Omega_d=\Omega^{blue}-\Omega^{red}<0\,,
\eqlabel{wn2}      
\end{equation}
Note that in this system, contrary to the standard ferromagnet \eqref{ferro}, the dimensionless temperature, defined by \eqref{t},
$t>0$ always. Let's denote by $\r_{11}^{c}$ the critical value of $\r_{11}$, \ie, 
\begin{equation}
\frac{m_b}{T}\bigg|_{\r_{11}=\r_{11}^{c}} =\frac{m_b}{T_c}\,.
\eqlabel{rcrit}
\end{equation}
Introducing 
\begin{equation}
\Delta\r_{11}=\r_{11}-\r_{11}^c\,,
\eqlabel{dr}
\end{equation}
from Fig.~\ref{figure1} it is clear that 
\begin{equation}
t\ \propto\  (\Delta\r_{11})^2\,,\qquad c_s^2\bigg|_{blue}\ \propto\  \left(-c_s^2\right)\bigg|_{red}\ \propto\  |\Delta\r_{11}|\ \propto\ t^{1/2}  \,.
\eqlabel{scaling}
\end{equation} 
Thus, 
\begin{equation}
c_\calh=-T \left(\frac{\del^2\calw}{\del T^2}\right)=\frac{s}{c_s^2}\bigg|_{red}^{blue}\propto c_s^{-2}\bigg|_{red}^{blue}
\propto t^{-1/2}\,,
\eqlabel{chn2}
\end{equation} 
where we used the fact that the entropy density $s$ is continuous across the phase transition. 
Comparing \eqref{chn2} and \eqref{defs0} we  determine the critical exponent $\a$ \cite{n2tr2}
\begin{equation}
\a=\frac 12\,.
\eqlabel{n2a}
\end{equation}
To determine the critical exponent $\b$ we need to identify the  control parameter corresponding to 
the external magnetic field $\calh$ of the effective ferromagnet of section 2. We propose to identify 
\begin{equation}
\calh=m_b\,.
\eqlabel{n2h}
\end{equation} 
Given \eqref{tcn2},  it follows from \eqref{scaling} that 
\begin{equation}
\del_\calh\ \propto\ -\del_t\ \propto\ -\frac{1}{\Delta\r_{11}}\ \del_{\Delta \r_{11}}\,.
\eqlabel{derh}
\end{equation}
The left plot on Fig.~\ref{figure2} implies that 
\begin{equation}
\calw=o(\Delta\r_{11}^2)\,,
\eqlabel{wfn2}
\end{equation}
at the very least. Actually, the best fit to $\Omega$ shows that\footnote{This is not surprising since $\Omega$ 
on the left plot of Fig.~\ref{figure2} can not be an even function of $\Delta\r_{11}$.} 
\begin{equation}
\calw\ \propto -|\Delta \r_{11}|^3\,.
\eqlabel{wfn21}
\end{equation}
Thus, using \eqref{derh} and \eqref{wfn21}, 
\begin{equation}
\calm=-\left(\frac{\del\calw}{\del \calh}\right)\ \propto\  \frac{1}{\Delta\r_{11}}\ \del_{\Delta \r_{11}}\calw\ \propto\ 
-|\Delta\r_{11}|\ \propto\ -t^{1/2}\,,
\eqlabel{mn2}
\end{equation}
Comparing \eqref{mn2} and \eqref{defs0} we conclude that 
\begin{equation}
\b=\frac 12\,.
\eqlabel{ben2}
\end{equation}
The rest of the critical exponents in \eqref{staticn2} follow from the scaling relations \eqref{screl}.

The second order phase transition in $\caln=2^*$ plasma we just exhibit is remarkably similar to the 
second order phase transition in $\caln=4$ supersymmetric Yang-Mills plasma with a single $U(1)\subset SU(4)_R$ 
R-symmetry chemical potential discussed in \cite{maeda}. In \cite{maeda} this chemical potential was identified 
with the external magnetic field $\calh$ of the effective ferromagnet of section 2.

\subsection{Cascading gauge theory plasma}

\begin{figure}[t]
\begin{center}
\psfrag{f}{{$\frac{32\pi^4}{81}\frac{\Omega}{\Lambda^4}$}}
\psfrag{TL}{{$\frac{T}{\Lambda}$}}
\psfrag{ks}{{$k_s$}}
  \includegraphics[width=3in]{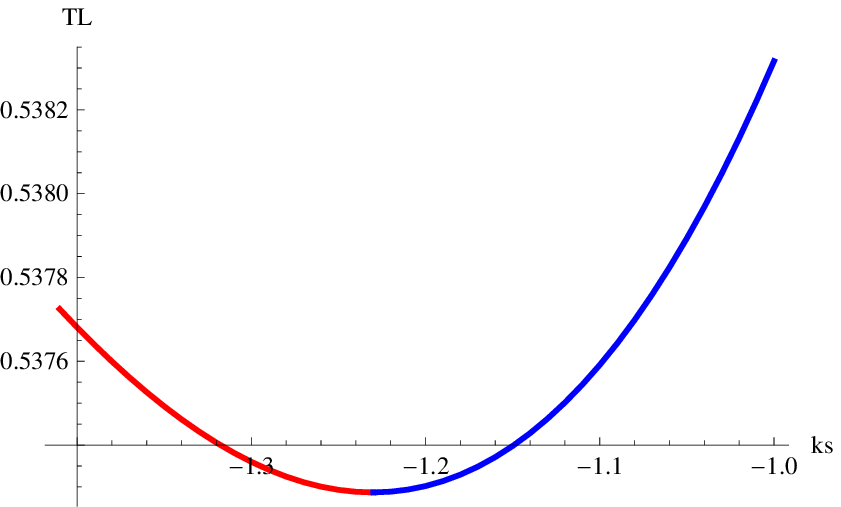}
\includegraphics[width=3in]{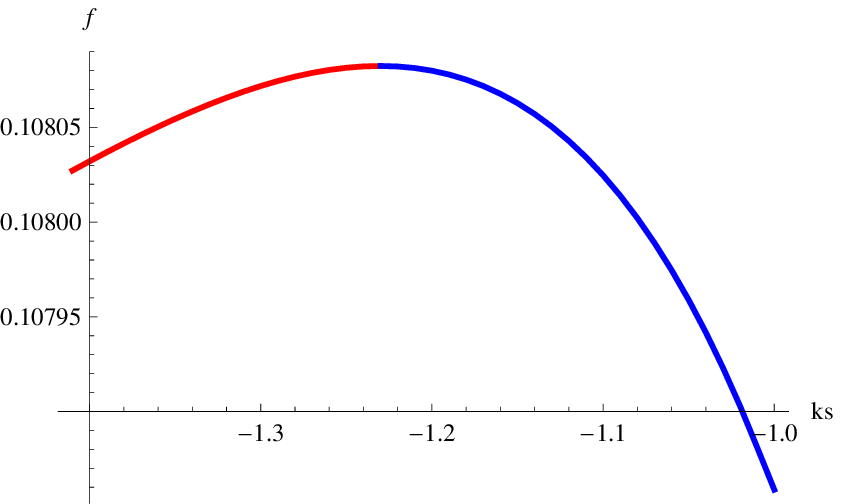}
\end{center}
  \caption{(Colour online)
The reduced temperature $\frac{T}{\Lambda}$ (left plot) and the 
free energy densities $\Omega_o$ of the ``ordered'' phase (blue curve, right plot) 
and $\Omega_d$ of the ``disordered'' phase (red curve, right plot) , see \eqref{diffw},
of the strongly coupled cascading plasma as a function of the dual gravitational parameter $k_s$.
} \label{figure3}
\end{figure}

The holographic renormalization and the thermodynamics of the cascading gauge theory plasma in the 
planar limit and at (infinitely) large 't Hooft coupling was discussed previously in 
\cite{ca3,ca4,ca5,cae1,ca6,castr1}. The cascading gauge theory \cite{ca2} is best thought 
of as $\caln=1$ supersymmetric $SU(K+P)\times SU(K)$ gauge theory where the effective number 
of colours $K$ is not constant along the renormalization group flow, but changes with energy 
according to \cite{ca3,cae2,cae3}
\begin{equation}
K=K(\mu)\sim 2P^2 \ln \frac{\mu}{\Lambda}\,,
\eqlabel{defk}
\end{equation}
at least when the energy scale $\mu$ is much larger than the strong coupling scale 
$\Lambda$ of the gauge theory. At zero temperature, and when $K(\mu)$ is a multiple of $P$,
cascading gauge theory confines in the infrared with the spontaneous breaking of chiral symmetry. 
The high temperature phase of the theory is expected to be that of the deconfined chirally symmetric plasma
\cite{ca3}. In was shown in \cite{ca6} that at  
\begin{equation}
T=T_{confinement}=0.6141111(3)\Lambda
\eqlabel{tde} 
\end{equation}
the deconfined chirally symmetric phase undergoes a first-order confinement phase transition, with spontaneous 
breaking of chiral symmetry. For  
\begin{equation}
T_c=0.8749(0)\times T_{confinement}\ < T\ < T_{confinement}\,,
\end{equation}
the deconfined phase of the cascading plasma, although non-perturbatively unstable due to the
nucleation of bubbles of the confined phase, remains perturbatively (and thermodynamically) 
stable \cite{castr1}. In the vicinity of $T_c$ the thermodynamics of the cascading gauge theory plasma is identical to that of the 
$\caln=2^*$ plasma. Fig.~\ref{figure3} presents the reduced temperature $\frac{T}{\Lambda}$ (left plot) 
and the free energy density of the cascading plasma as a function of the 
dual gravitational parameter $k_s$.  We use the same colour coding as in Fig.~\ref{figure1} and Fig.~\ref{figure2}:
the ``blue'' dots indicate stable (ordered) phase and the "red'' dots indicate unstable (disordered) phase. 
As in the case of $\caln=2^*$ plasma, in the ordered (disordered) phase the square of the speed of sound is positive (negative)
\cite{castr1}.  Once we identify the external magnetic field $\calh$ of the effective ferromagnet of section 2 as
\begin{equation}
\calh=\Lambda\,,
\eqlabel{cash}
\end{equation}
we can literally repeat the arguments of section 3.1 and arrive at the same set of static critical exponents 
\eqref{staticn2}.

\subsection{"Exotic black holes''}

\begin{figure}[t]
\begin{center}
\psfrag{Fred}{{$\frac{\Omega}{(\pi T)^3}$}}
\psfrag{TcT}{{$\frac{T_c}{T}$}}
\psfrag{O}{{$\langle\calo_i\rangle>^2$}}
\includegraphics[width=3in]{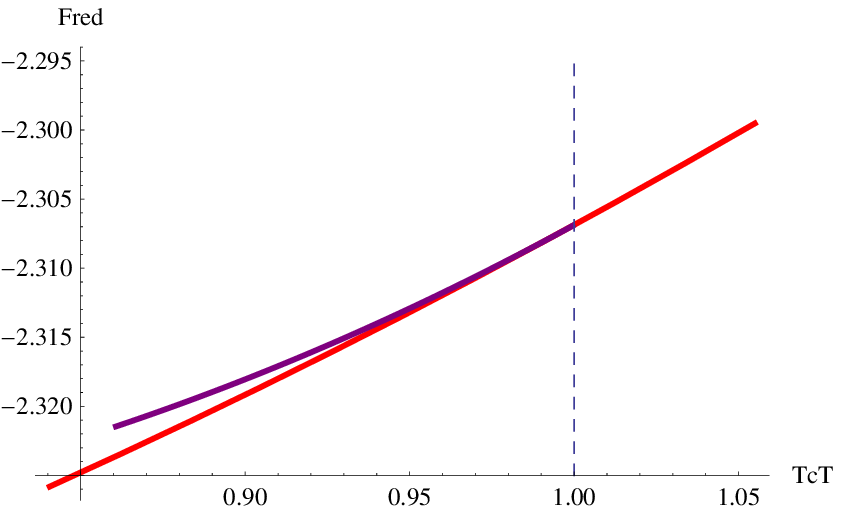}
\includegraphics[width=3in]{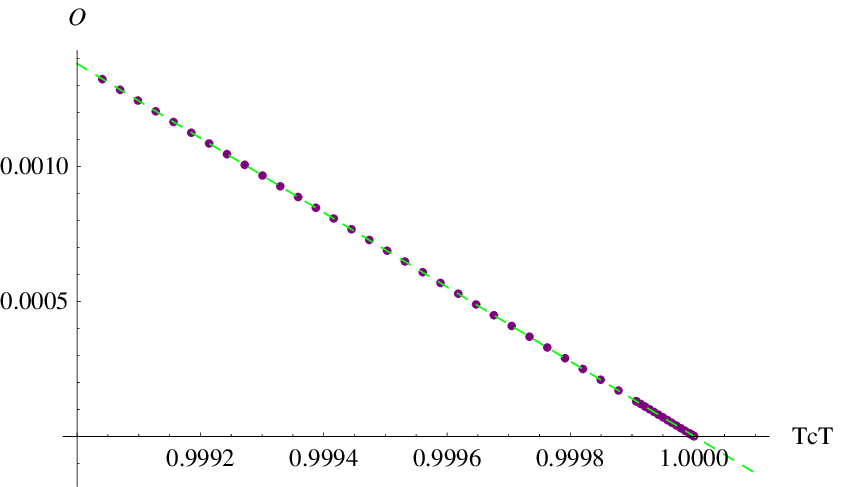}
\end{center}
  \caption{(Colour online)
The free energy densities  $\Omega_0$ of the ``ordered'' phase (red curve, left plot) and $\Omega_d$ of the ``disordered'' 
phase (purple curve, left plot) as a function of the reduced temperature $\frac{T}{T_c}$ in gauge theory plasma dual to holographic RG 
flow in \cite{bp1}. The right plot represents the square of $\langle\calo_i\rangle$  (which we use an an order parameter for the transition) as a function 
of reduced temperature. The dashed 
green line is a linear fit to  $\langle\calo_i\rangle^2$.
 } \label{figure4}
\end{figure}

\begin{figure}[t]
\begin{center}
\psfrag{cs}{{$2 c_s^2$}}
\psfrag{tt}{{$\frac{T_c}{T}$}}
\includegraphics[width=3in]{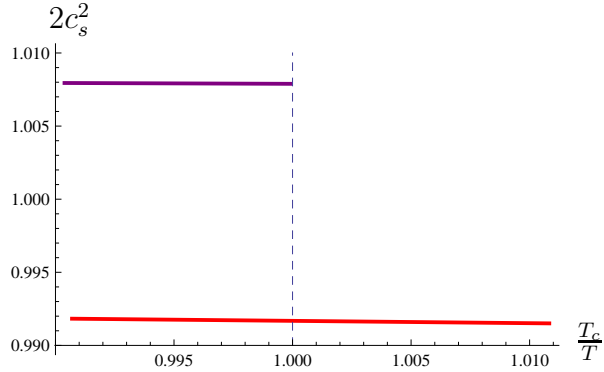}
\end{center}
  \caption{(Colour online)
The speed of sound $c_s$ in gauge theory plasma dual to holographic RG 
flow in \cite{bp1} as a function of the reduced temperature $\frac{T_c}{T}$.
 }\label{figure5}
\end{figure}

Following Gubser's suggestion \cite{g}, we constructed an exotic model of the second order phase transition in $d=3$ at finite temperature 
and zero chemical potentials in \cite{bp1}. Specifically, we considered relativistic conformal field theory in 
2+1 dimensions, deformed by a relevant operator $\calo_r$:
\begin{equation}
H_{CFT}\to \tilde{H}=H_{CFT}+\lambda_r \calo_r\,.
\eqlabel{ham3d}
\end{equation}
Such a deformation softly breaks the scale invariance and induces the renormalization group flow. 
We further assumed that the deformed theory $\tilde{H}$ has an irrelevant operator $\calo_i$ that mixes along the 
RG flow with $\calo_r$. In the explicit holographic model realizing this scenario \cite{bp1}, the irrelevant operator 
$\calo_i$ developed a vacuum expectation value in the high temperature phase, \ie, for $T>T_c$, spontaneously breaking a discrete $\zet_2$ symmetry of the model.
The unusual part of this phase transition was that the symmetry broken phase occurs at high temperatures (rather then the low temperatures) and that the 
broken phase (although perturbatively and thermodynamically stable) has higher free energy density than the unbroken phase with 
$\langle\calo_i\rangle=0$. The various high temperature phases which spontaneously break $\zet_2$ symmetry differ in the holographic 
wavefunction of the condensate and their free energies \cite{bp1}.  Without the loss of generality we focus here on the 
model with $\dim[\calo_r]=2$, and the broken phase of the lowest free energy.  Fig.~\ref{figure4} represents the energy densities 
of the ordered (red line, left plot) and the disordered (purple line, left plot) phases as a function of the reduced temperature $\frac{T}{T_c}$.
The expectation value of $\calo_i$ is the order parameter of the phase transition --- the spontaneous magnetization 
$\calm$ of the effective ferromagnet in section 2.  An excellent linear fit (the dashed green line) 
to $\langle\calo_i\rangle^2$ (see the right plot on Fig.~\ref{figure4}) implies 
that in this model 
\begin{equation}
\calm\equiv \langle\calo_i\rangle\ \propto |t|^{1/2}\,,
\eqlabel{b3d}
\end{equation}
resulting in the critical exponent 
\begin{equation}
\b=\frac 12\,.
\eqlabel{beta3d}
\end{equation}
From Fig.~\ref{figure5} it is clear that the speed of sound is finite at the transition\footnote{Notice that 
in the disordered phase the speed of sound violates the bound proposed in \cite{v1,v2}.}, thus 
\begin{equation}
c_\calh\ \propto c_s^{-2}\ \propto |t|^0\,,
\eqlabel{ch3d}
\end{equation}
resulting in the critical exponent 
\begin{equation}
\a=0\,.
\eqlabel{alpha3d}
\end{equation}
The remaining critical exponents can be obtained from the scaling relations \eqref{screl}:
\begin{equation}
\left(\a,\b,\gamma,\delta,\nu,\eta\right)=\left(0,\frac 12,1, 3,1, 1 \right)\,.
\eqlabel{static3d}
\end{equation}
Note that the universality classes \eqref{staticn2} and \eqref{static3d} are different.

\section{Holographic bulk viscosity at criticality}
The technique for computing the bulk viscosity in non-conformal holographic models from the dispersion relation of the sound waves was developed
in \cite{bbs}\footnote{A much simpler computational framework  was recently proposed in 
\cite{s1,s2}.}. It was used to compute the bulk viscosity of $\caln=2^*$ plasma in \cite{bbs,n2tr2,bp2}, and the bulk viscosity of the cascading 
gauge theory plasma  in \cite{castr0,castr1}. It is straightforward to generalize the method of \cite{bbs} to the computation of the bulk viscosity of 
the gauge theory plasma dual to the holographic RG flow \cite{bp1}. The details of the latter analysis will appear elsewhere \cite{pph},
and here we report only the results.

In the rest of this section for each of the holographic models with a second order phase transition discussed above we analyze the scaling of the 
bulk viscosity at criticality in the framework of Models A,B,C ---  \eqref{kkts}-\eqref{on}. 

\subsection{$N=2^*$ plasma }
The main result of \cite{n2tr2,bp2} was that the ratio of the bulk-to-shear  viscosities in $\caln=2^*$ plasma remains finite 
across the second order phase transition\footnote{The second-order transport coefficient is nonetheless divergent \cite{rel}.}:
\begin{equation}
\frac{\zeta}{\eta}\ \propto |t|^0\,.
\eqlabel{bulkn2}
\end{equation} 
Since in holographic phase transitions under the consideration both the entropy density and the shear viscosity are finite, 
we conclude from \eqref{bulkn2} that the bulk viscosity stays finite. Thus:
\nxt Model A is inconsistent with holographic analysis as it predicts divergent bulk viscosity, $\zeta\ \propto\ |t|^{-1/2}$\ ;
\nxt Model B does not contradict our holographic analysis as it predicts that $\zeta_{singular}\ \propto\ |t|^{3/2} $\ ;
\nxt Model C agrees with holographic analysis provided the dynamical exponent $z$ is 
\begin{equation}
z=1\,.
\eqlabel{zn2}
\end{equation}

\subsection{Cascading gauge theory plasma}
It was argued in \cite{castr1} that the bulk viscosity of the cascading gauge theory plasma remains finite across the second order phase 
transition much like in the case of $\caln=2^*$ plasma. Thus, cascading gauge theory plasma appears to share not only the static universality 
class with $\caln=2^*$ model, but also the dynamical one --- in particular, the critical exponent $z$ is given by \eqref{zn2}.

\subsection{"Exotic black holes''}
\begin{figure}[t]
\begin{center}
\psfrag{zetaeta}{{$\frac{\zeta}{\eta}$}}
\psfrag{lnzetaeta}{{$\ln\left(\frac{\zeta}{\eta}\right)$}}
\psfrag{cs}{{$\left(\frac 12-c_s^2\right)$}}
\psfrag{lnchi4}{{$\ln|\langle\calo_i\rangle|$}}
\includegraphics[width=3in]{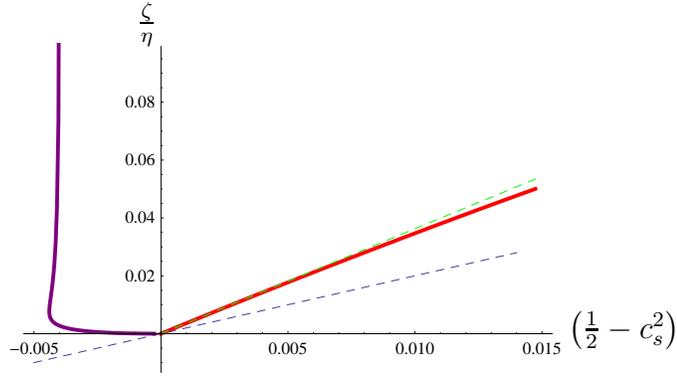}
\end{center}
  \caption{(Colour online)
The ratio of bulk-to-shear viscosities in gauge theory plasma dual to the RG flow in \cite{bp1}.
The red curve corresponds to the ``ordered'' phase of the theory, and the purple one to the ``disordered'' phase.
The dashed blue line indicates the bulk viscosity bound proposed in \cite{n2tr2}. The dashed green line 
corresponds to the high-temperature approximation to the viscosity ratio given by \eqref{hight}. 
 } \label{figure6}
\end{figure}

\begin{figure}[t]
\begin{center}
\psfrag{zetaeta}{{$\frac{\zeta}{\eta}$}}
\psfrag{lnzetaeta}{{$\ln\left(\frac{\zeta}{\eta}\right)$}}
\psfrag{cs}{{$\left(\frac 12-c_s^2\right)$}}
\psfrag{lnchi4}{{$\ln|\langle\calo_i\rangle|$}}
\includegraphics[width=3in]{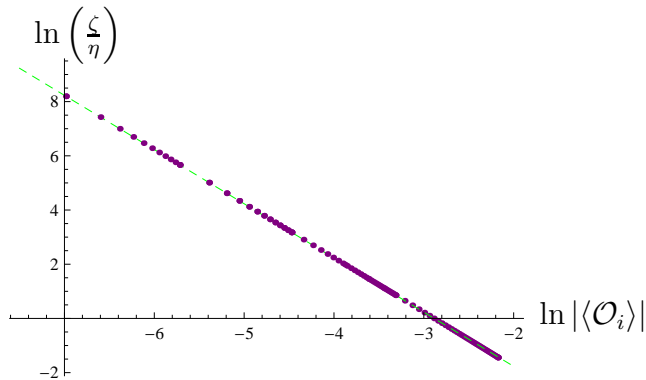}
\end{center}
  \caption{(Colour online)
The ratio of bulk-to-shear viscosities in gauge theory plasma dual to the RG flow in \cite{bp1} in the vicinity 
of the second order phase transition as a function of the order parameter. The dashed green line represents the 
linear fit to the log-log data plot with the slope of $(-1.9999(6))$.  
 } \label{figure7}
\end{figure}

Fig.~\ref{figure6} represents the ratio of the bulk-to-shear viscosities of the gauge theory plasma dual to the 
RG flow in \cite{bp1}. The red curve corresponds to the "ordered'' (symmetric) phase that the purple curve corresponds to
the ``disordered'' (broken) phase. The dashed blue line represents the bulk viscosity bound 
\begin{equation}
\frac{\zeta}{\eta}\ge 2 \left(\frac 12-c_s^2\right)\,,
\eqlabel{bound}
\end{equation}  
proposed in \cite{n2tr2}. The dashed green line represents the bulk-to-shear viscosity ratio for a symmetric phase at 
high temperatures:
\begin{equation}
\frac{\zeta}{\eta}\bigg|_{ordered}=\frac{2\pi}{\sqrt{3}}\left(\frac 12 -c_s^2\right)+\calo\left(\left(\frac12-c_s^2\right)^2\right)\,.
\eqlabel{hight}
\end{equation} 
Notice that the bulk viscosity bound is satisfied both in the symmetric and in the broken phases. 
In fact, in the disorder phase (purple curve) the bound is satisfied trivially as here $c_s^2>\frac 12$, 
see also  Fig.~\ref{figure5}.

Fig.~\ref{figure6} shows a rapid rise of the bulk-to-shear viscosity ratio in the disordered (purple curve) phase as one approaches the transition. 
In fact, a detailed analysis presented in Fig.~\ref{figure7} strongly suggests that bulk viscosity diverges in this case. There, we plot 
$\ln\left(\frac{\zeta}{\eta}\right)$ versus  $\ln|\langle\calo_i\rangle|$. A dashed green line given by 
\begin{equation}
y=-1.9999(6)\ x-5.7684(5)
\eqlabel{green}
\end{equation}
provides an excellent linear fit to the data.
Thus, we expect 
\begin{equation}
\frac{\zeta}{\eta}\bigg|_{disordered}\ \propto\ |\langle\calo_i\rangle|^{-2}\ \propto |t|^{-1}\,,
\eqlabel{zeta3d}
\end{equation}
where we used \eqref{b3d}.
From \eqref{zeta3d} we conclude:
\nxt once again, Model A is inconsistent with our holographic analysis as it predicts a finite bulk viscosity at the transition, 
$\zeta\ \propto |t|^0$\ ;
\nxt Model B is inconsistent as well, as it predicts $\zeta_{singular}\ \propto\ |t|^1$\ ; 
\nxt Model C agrees with holographic analysis provided the dynamical exponent $z$ is 
\begin{equation}
z=1\,.
\eqlabel{z3d}
\end{equation}
Rather intriguingly, even though the static universality classes of the $\caln=2^*$ (and the cascading) gauge theory plasma
and the exotic plasma discussed in this section are different, when the bulk viscosity is interpreted in the 
framework of Model C, both classes  have the same dynamical critical exponent $z$.

\section{Conclusion}
In this paper we studied bulk viscosity at criticality of the second order phase transition in non-conformal models 
of gauge theory/string theory correspondence at finite temperature, but at zero chemical potential for the conserved charges.
We identified the second order phase transitions in $\caln=2^*$ and the cascading gauge theory plasma and showed that the two models 
are in the same static universality class. Somewhat surprisingly, this universality class is {\it not} of the mean-field theory type.
Incidentally, the same universality class contains a second order phase transition in 
$\caln=4$ supersymmetric Yang-Mills plasma with a single $U(1)\subset SU(4)_R$ R-symmetry chemical potential discussed in \cite{maeda}.  
The static universality class of the $\caln=2^*$ plasma is different from the static universality class of the gauge theory 
dual to the 3+1 dimensional asymptotically $AdS_4$ RG flow discussed in \cite{bp1}. Having different static universality classes in our holographic laboratory 
allowed us to test different proposals for the scaling of bulk viscosity in the vicinity of the second order phase transition. 
We found that neither the model of \cite{bulk1} or \cite{bulk2} conforms to the tests. If dynamical scaling of bulk viscosity proposed by Onuki \cite{bulk3} 
is correct, then all our holographic models must have the same dynamical critical exponent $z=1$.    
  
In the future, it is important to explicitly verify the scaling relations \eqref{screl} in our holographic models. 
It would be extremely interesting to subject Onuki's theory \cite{bulk3} to a direct holographic test by independently computing the 
dynamical critical exponent $z$ from the scaling relation \eqref{taus}. Finally, it is interesting to understand the dynamical universality 
class of the holographic models. We already know that in such theories the ratio of shear viscosity to the entropy density is universal. 
Is it possible that the universality of the holographic transport at critically is more robust?

\section*{Acknowledgments}
We would like to thank Pavel Kovtun for valuable discussions and correspondence.
Research at Perimeter Institute is
supported by the Government of Canada through Industry Canada and by
the Province of Ontario through the Ministry of Research \&
Innovation. AB gratefully acknowledges further support by an NSERC
Discovery grant and support through the Early Researcher Award
program by the Province of Ontario.


\begin{thebibliography}{99}


\bibitem{m9711}
J.~M.~Maldacena,
``The large $N$ limit of superconformal field theories and supergravity,''
Adv.\ Theor.\ Math.\ Phys.\  {\bf 2}, 231 (1998)
[Int.\ J.\ Theor.\ Phys.\  {\bf 38}, 1113 (1999)]
[arXiv:hep-th/9711200].


\bibitem{bpta} A.~Buchel and C.~Pagnutti, to appear.

\bibitem{kovs}
  P.~K.~Kovtun and A.~O.~Starinets,
  Phys.\ Rev.\  D {\bf 72}, 086009 (2005)
  [arXiv:hep-th/0506184].

\bibitem{kh} A.~Kogan and H.~Meyer, J.\ Low Temp.\ Phys.\ {\bf 110}, 899 (1998)



\bibitem{u1}
  A.~Buchel and J.~T.~Liu,
  Phys.\ Rev.\ Lett.\  {\bf 93}, 090602 (2004)
  [arXiv:hep-th/0311175].

\bibitem{u2}
  P.~Kovtun, D.~T.~Son and A.~O.~Starinets,
  Phys.\ Rev.\ Lett.\  {\bf 94}, 111601 (2005)
  [arXiv:hep-th/0405231].

\bibitem{u3}
  A.~Buchel,
  Phys.\ Lett.\  B {\bf 609}, 392 (2005)
  [arXiv:hep-th/0408095].




\bibitem{bbs}
  P.~Benincasa, A.~Buchel and A.~O.~Starinets,
  Nucl.\ Phys.\  B {\bf 733}, 160 (2006)
  [arXiv:hep-th/0507026].


\bibitem{bulk1}
  F.~Karsch, D.~Kharzeev and K.~Tuchin,
  Phys.\ Lett.\  B {\bf 663}, 217 (2008)
  [arXiv:0711.0914 [hep-ph]].

\bibitem{bulk2}
  C.~Sasaki and K.~Redlich,
  Phys.\ Rev.\  C {\bf 79}, 055207 (2009)
  [arXiv:0806.4745 [hep-ph]].

\bibitem{bulk3}
A.~Onuki,    Phys.\ Rev.\  E {\bf 55} 403  (1997).


\bibitem{n2tr1}
  A.~Buchel, S.~Deakin, P.~Kerner and J.~T.~Liu,
  Nucl.\ Phys.\  B {\bf 784}, 72 (2007)
  [arXiv:hep-th/0701142].

\bibitem{n2tr2}
  A.~Buchel,
  Phys.\ Lett.\  B {\bf 663}, 286 (2008)
  [arXiv:0708.3459 [hep-th]].

\bibitem{castr1}
  A.~Buchel,
  Nucl.\ Phys.\  B {\bf 820}, 385 (2009)
  [arXiv:0903.3605 [hep-th]].

\bibitem{pw1}
  K.~Pilch and N.~P.~Warner,
  Nucl.\ Phys.\  B {\bf 594}, 209 (2001)
  [arXiv:hep-th/0004063].



\bibitem{pw2}
  A.~Buchel, A.~W.~Peet and J.~Polchinski,
  Phys.\ Rev.\  D {\bf 63}, 044009 (2001)
  [arXiv:hep-th/0008076].

\bibitem{pw3}
  N.~J.~Evans, C.~V.~Johnson and M.~Petrini,
  JHEP {\bf 0010}, 022 (2000)
  [arXiv:hep-th/0008081].



\bibitem{ca1}
  I.~R.~Klebanov and A.~A.~Tseytlin,
  Nucl.\ Phys.\  B {\bf 578}, 123 (2000)
  [arXiv:hep-th/0002159].


\bibitem{ca2}
  I.~R.~Klebanov and M.~J.~Strassler,
  JHEP {\bf 0008}, 052 (2000)
  [arXiv:hep-th/0007191].


\bibitem{ca3}
  A.~Buchel,
  Nucl.\ Phys.\  B {\bf 600}, 219 (2001)
  [arXiv:hep-th/0011146].




\bibitem{ca4}
  A.~Buchel, C.~P.~Herzog, I.~R.~Klebanov, L.~A.~Pando Zayas and A.~A.~Tseytlin,
  JHEP {\bf 0104}, 033 (2001)
  [arXiv:hep-th/0102105].

\bibitem{ca5}
  S.~S.~Gubser, C.~P.~Herzog, I.~R.~Klebanov and A.~A.~Tseytlin,
  JHEP {\bf 0105}, 028 (2001)
  [arXiv:hep-th/0102172].

\bibitem{ca6}
  O.~Aharony, A.~Buchel and P.~Kerner,
  Phys.\ Rev.\  D {\bf 76}, 086005 (2007)
  [arXiv:0706.1768 [hep-th]].


\bibitem{bp1}
  A.~Buchel and C.~Pagnutti,
  Nucl.\ Phys.\  B {\bf 824}, 85 (2010)
  [arXiv:0904.1716 [hep-th]].

\bibitem{bp2}
  A.~Buchel and C.~Pagnutti,
  Nucl.\ Phys.\  B {\bf 816}, 62 (2009)
  [arXiv:0812.3623 [hep-th]].

\bibitem{castr0}
  A.~Buchel,
  Phys.\ Rev.\  D {\bf 72}, 106002 (2005)
  [arXiv:hep-th/0509083].


\bibitem{hh}
  P.~C.~Hohenberg and B.~I.~Halperin,
  Rev.\ Mod.\ Phys.\  {\bf 49}, 435 (1977).

\bibitem{hhh}
  S.~A.~Hartnoll, C.~P.~Herzog and G.~T.~Horowitz,
  Phys.\ Rev.\ Lett.\  {\bf 101}, 031601 (2008)
  [arXiv:0803.3295 [hep-th]].

\bibitem{maeda}
  K.~Maeda, M.~Natsuume and T.~Okamura,
  Phys.\ Rev.\  D {\bf 78}, 106007 (2008)
  [arXiv:0809.4074 [hep-th]].

\bibitem{n2e1}
  A.~Buchel and J.~T.~Liu,
  JHEP {\bf 0311}, 031 (2003)
  [arXiv:hep-th/0305064].

\bibitem{n2e2}
  A.~Buchel,
  Nucl.\ Phys.\  B {\bf 708}, 451 (2005)
  [arXiv:hep-th/0406200].

\bibitem{larry}
  S.~Paik and L.~G.~Yaffe,
  ``Thermodynamics of SU(2) N=2 supersymmetric Yang-Mills theory,''
  arXiv:0911.1392 [hep-th].

\bibitem{cae1}
  O.~Aharony, A.~Buchel and A.~Yarom,
  Phys.\ Rev.\  D {\bf 72}, 066003 (2005)
  [arXiv:hep-th/0506002].

\bibitem{cae2}
  M.~Krasnitz,
  JHEP {\bf 0212}, 048 (2002)
  [arXiv:hep-th/0209163].

\bibitem{cae3}
  O.~Aharony, A.~Buchel and A.~Yarom,
  JHEP {\bf 0611}, 069 (2006)
  [arXiv:hep-th/0608209].

\bibitem{g}
  S.~S.~Gubser,
  Class.\ Quant.\ Grav.\  {\bf 22}, 5121 (2005)
  [arXiv:hep-th/0505189].

\bibitem{v1}
  P.~M.~Hohler and M.~A.~Stephanov,
  Phys.\ Rev.\  D {\bf 80}, 066002 (2009)
  [arXiv:0905.0900 [hep-th]].

\bibitem{v2}
  A.~Cherman, T.~D.~Cohen and A.~Nellore,
  Phys.\ Rev.\  D {\bf 80}, 066003 (2009)
  [arXiv:0905.0903 [hep-th]].


\bibitem{s1}
  A.~Cherman and A.~Nellore,
  Phys.\ Rev.\  D {\bf 80}, 066006 (2009)
  [arXiv:0905.2969 [hep-th]].


\bibitem{s2}
  A.~Yarom,
  ``Notes on the bulk viscosity of holographic gauge theory plasmas,''
  arXiv:0912.2100 [hep-th].



\bibitem{pph}
C.~Pagnutti, Ph.D. thesis.

\bibitem{rel}
  A.~Buchel,
  Phys.\ Lett.\  B {\bf 681}, 200 (2009)
  [arXiv:0908.0108 [hep-th]].



\end{thebibliography}
\end{document}